\documentclass{aa}
\usepackage{natbib}
\bibpunct{(}{)}{;}{a}{}{,} % to follow the A&A style
\usepackage{txfonts}
\usepackage[colorlinks=true,allcolors=blue]{hyperref}
\usepackage{upgreek}
\usepackage{newtxtext,newtxmath}

\usepackage[T1]{fontenc}

\usepackage{graphicx}   % Including figure files
\usepackage{amsmath}    % Advanced maths commands
\usepackage{amssymb}    % Extra maths symbols
\usepackage[normalem]{ulem}
\usepackage{caption}
\usepackage{subcaption}
\usepackage{multirow}
\usepackage{adjustbox} % Limit table
\usepackage{lscape}  %landscape formatting
\usepackage{array}
\usepackage{float}
\usepackage{orcidlink}

\newcolumntype{C}[1]{>{\centering\arraybackslash}p{#1}}

\begin{document}

\title{Complex angular structure of three elliptical galaxies from high-resolution ALMA observations of strong gravitational lenses}
\titlerunning{Angular structure in ALMA observations of strong lenses}

\author{
\orcidlink{0000-0002-8999-9636}H.\,R.~Stacey\inst{1,2}\thanks{E-mail: hannah.stacey@eso.org},
\orcidlink{0000-0002-4912-9943}D.\,M.~Powell\inst{2},
S.~Vegetti,\inst{2}
\orcidlink{0000-0003-1787-9552}J.\,P.~McKean\inst{3,4,5},
\orcidlink{0000-0002-4030-5461}C.\,D.~Fassnacht\inst{6},
\orcidlink{0000-0003-2812-8607}D.\,Wen\inst{3}
\and
\orcidlink{0000-0003-2227-1998}C.\,M.~O'Riordan\inst{2}
}
\authorrunning{H.\,R.\,Stacey et al.}

\institute{
% List of institutions
European Southern Observatory (ESO), Karl-Schwarzschild Str. 2, D-85748 Garching bei M\"unchen, Germany 
\and
Max Planck Institute for Astrophysics, Karl-Schwarzschild Str. 1, D-85748 Garching bei M\"unchen, Germany
\and
Kapteyn Astronomical Institute, University of Groningen, Postbus 800, NL-9700 AV Groningen, The Netherlands
\and
South African Radio Astronomy Observatory (SARAO), P.O. Box 443, Krugersdorp 1740, South Africa
\and
Department of Physics, University of Pretoria, Lynnwood Road, Hatfield, Pretoria, 0083, South Africa
\and
Department of Physics and Astronomy, University of California Davis, 1 Shields Avenue, Davis, CA 95616, USA
}

% These dates will be filled out by the publisher
\date{Received XXX; accepted YYY}

% Abstract of the paper
\abstract{
The large-scale mass distributions of galaxy-scale strong lenses have long been assumed to be well described by a singular ellipsoidal power-law density profile with external shear. However, the inflexibility of this model could lead to systematic errors in astrophysical parameters inferred with gravitational lensing observables. Here, we present observations with the Atacama Large (sub-)Millimetre Array (ALMA) of three strongly lensed dusty star-forming galaxies at $\simeq30$~mas angular resolution and investigate the sensitivity of these data to angular structure in the lensing galaxies. We jointly infer the lensing mass distribution and the full surface brightness of the lensed sources with multipole expansions of the power-law density profile up to the fourth order using a technique developed for interferometric data. All three datasets strongly favour third and fourth-order multipole amplitudes of $\approx1$~percent of the convergence. While the infrared stellar isophotes and isodensity shapes agree for one lens system, for the other two the isophotes disagree to varying extents, suggesting contributions to the angular structure from dark matter intrinsic or extrinsic to the lensing galaxy.
}

\keywords{Gravitational lensing: strong -- Submillimeter: general -- Galaxies: elliptical and lenticular, cD}
            
\maketitle

%%%%%%%%%%%%%%%%%%%%%%%%%%%%%%%%%%%%%%%%%%%%%%%%%%

%%%%%%%%%%%%%%%%% BODY OF PAPER %%%%%%%%%%%%%%%%%%

\section{Introduction}

Strong gravitational lensing occurs when the light from a distant background galaxy is distorted and magnified into several distinct images by a massive foreground galaxy. This phenomenon has become a well-established tool with which to probe the mass structure and dark matter distribution within the lens galaxies. For example, the observed surface brightness distribution of the lensed images has been used to constrain the evolutionary pathways of elliptical galaxies \citep{Sonnenfeld:2012}, cosmological parameters (e.g. \citealt{Wong:2020,Collett:2018}) and the nature of dark matter via the gravitational effect of low-mass dark matter halos \citep{Vegetti:2014,Vegetti:2023,Ritondale:2019a,Hsueh:2020,Gilman:2020a,Enzi:2021,Powell:2023}. Moreover, the magnification provided by the lens increases the effective angular resolution of the data, allowing the detailed astrophysical processes within extremely distant objects at important cosmological epochs to be studied (e.g. \citealt{Yang:2019,Rizzo:2021,Stacey:2022,Geach:2023}). Fundamental to all these important lines of research with strong gravitational lensing is the parameterisation of the large-scale mass distribution of the lensing galaxy (the so-called `macro model'). For galaxy-scale strong lensing events, this is commonly assumed to be described by an ellipsoidal power-law mass-density profile with some external shear component to account for the sometimes complex environment of the lens. This simple parameterisation can lead to valid questions regarding the accuracy of the lens model or the robustness of the reconstructed source, given the uncertainties in the lens modelling.

Indeed, several recent works have highlighted the limitations of the power-law model assumption for the mass distribution of the lens and the impact this can have on the astrophysical applications of galaxy-scale gravitational lensing \citep{Powell:2021,Cao:2022,VandeVyvere:2022,Nightingale:2022,He:2023,O'Riordan:2023}. The majority of elliptical galaxies have boxy or discy isophote shapes in their stellar distribution \citep{Bender:1988,Bender:1989,Cappellari:2016} and can exhibit twists and shape variations with radius \citep{Liller:1960,Liller:1966,King:1978}. These features should manifest an anisotropic density structure that may result in systematic errors in the recovered Hubble constant from lensing \citep{Cao:2022,VandeVyvere:2022}, change the relative predicted image magnifications by $10\--40$~percent \citep{Evans:2003,Congdon:2005,Powell:2022,Cohen:2024}, and induce false detections of low-mass dark matter haloes \citep{Nightingale:2022,He:2023,O'Riordan:2023,Cohen:2024}, leading to biased constraints on dark matter models. 

\begin{table*}[!h]
    \caption{Data used in the analysis. For each lens system, we give the lens redshift ($z_l$), source redshift ($z_s$), naturally weighted synthesised beam FWHM (major and minor axis), central frequency ($\nu$), total continuum bandwidth ($\Delta\nu$), channel width ($\delta\nu$), RA and Dec of the phase centre (J2000), and ALMA project code.}
    \centering
    \adjustbox{width=\textwidth}{
    \begin{tabular}{l c c c c c c c c l} \hline \rule{0pt}{2ex}
    Name & $z_l$ & $z_s$ & FWHM & $\nu$ & $\Delta\nu$ & $\delta\nu$ & RA & Dec & Project code \\ 
     &  & & (arcsec) & (GHz) & (GHz) & (MHz) & (deg) & (deg) &  \\ \hline \rule{0pt}{2.2ex}
    SDP\,81 & 0.30 & 3.04 & $0.024\times0.033$ & 290
    & 8 & 500 & 135.7983748506 & +0.651860959014 & 2011.0.00016.SV \\ 
    SPT\,0532$-$50    & 1.15 & 3.40 &  $0.035\times0.050$ & 352
    & 4 & 62.5 & 83.21270432730 & $-$50.78545546562 & 2016.1.01374.S \\
    SPT\,0538$-$50    & 0.40 & 2.78 &  $0.026\times0.029$ & 347
    & 8 & 62.5 & 84.57012499581 & $-$50.51444470965 & 2016.1.01374.S \\ \hline
    \end{tabular}
    }
    \label{table:obs}
\end{table*}

Recently, \cite{Powell:2022} showed that complex angular structure in the lens and an external shear gradient were required to focus the source with observations at milli-arcsecond (mas) resolution (see also \citealt{Spingola:2018}), suggesting that the sensitivity to angular structure increases with angular resolution. In this regard, observations of bright and highly magnified starburst galaxies at high redshift with the Atacama Large Millimetre/sub-millimetre Array (ALMA) can provide a detailed test of angular structure, given the higher angular resolution when compared to current optical or infrared telescopes. Here, we have analysed publicly available data from the ALMA archive for three gravitational lenses observed at an angular resolution of $\approx30$~mas. By using a pixellated lens modelling methodology adapted for interferometric data \citep{Vegetti:2009,Rybak:2015a,Rybak:2015b,Rizzo:2018,Ritondale:2019a,Powell:2021,Powell:2022}, we investigated the angular structure in these lens galaxies with multipole expansions of the power-law density profile. 

The paper is structured as follows. In Section~\ref{section:data} we give an overview of the ALMA observations and data reduction. This section also includes a review of available {\it Hubble} Space Telescope (HST) infrared imaging, which we use to compare the mass and light distributions of the three lens galaxies. In Section~\ref{section:modelling}, we describe the lens model parameterisations and methodology. In Section~\ref{section:results}, we report our results and in Section \ref{section:discussion}, we discuss the possible origins of the angular structure, before summarising and considering the direction of future work in Section~\ref{section:conclusions}.

\begin{figure*}
    \centering
    \includegraphics[height=0.19\textheight]{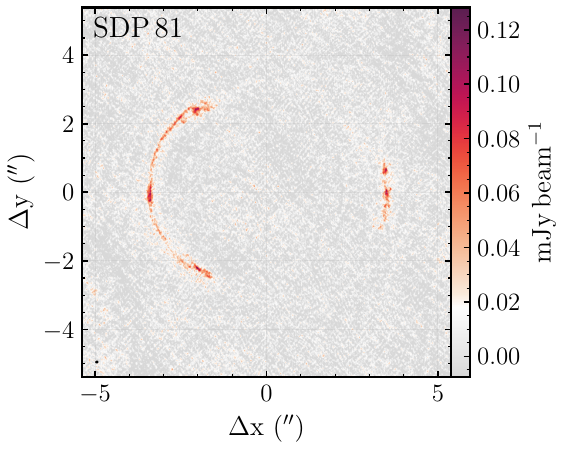}\hspace{5pt}
    \includegraphics[height=0.19\textheight]{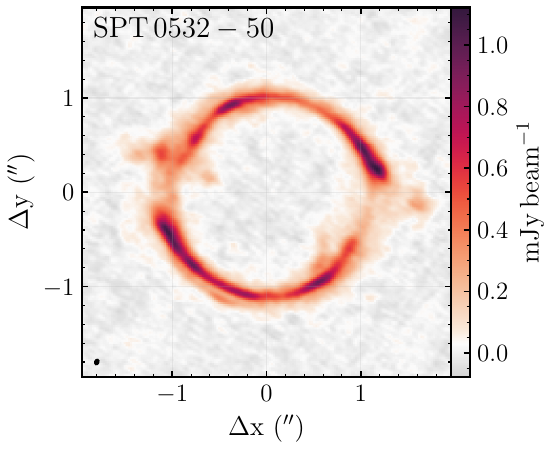}\hspace{5pt}
    \includegraphics[height=0.19\textheight]{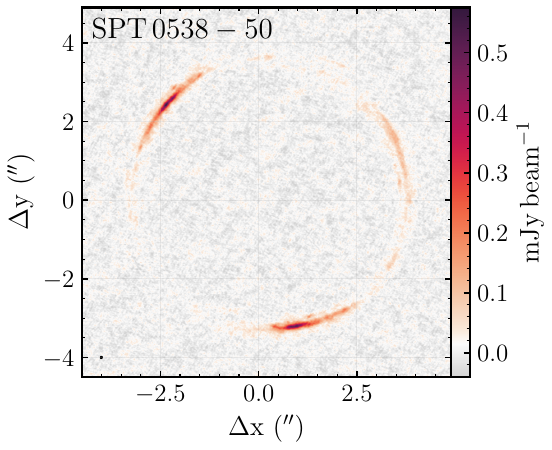}
    \caption{Deconvolved images using natural weighting of the visibilities. The synthesised beam is shown by the black ellipse in the bottom left corner of each panel and the sizes are given in Table~\ref{table:obs}. These images are only for visualisation as the lens modelling was done with respect to the visibility data directly. 
    %The contours show a signal-to-noise ratio of 3, 6, 12, 24, 48.  
    }
    \label{fig:clean_images}
\end{figure*}

\section{Data reduction}
\label{section:data}

\subsection{ALMA observations}

We obtained data from the ALMA archive for three gravitationally lensed dusty star-forming galaxies: SPT\,0532$-$50 (SPT\,S\,J053250$-$5047.1), SPT\,0538$-$50 (SPT\,S\,J053816$-$5030.8), and SDP\,81 (H-ATLAS\,J090311.6$+$003906). These are galaxy-scale lenses with known source and lens redshifts (\citealt{Negrello:2010,Spilker:2016}; Table~\ref{table:obs}). Details of the lens systems and ALMA observations are given in Table~\ref{table:obs}. The observations of SPT\,0532$-$50 and SPT\,0538$-$50 consisted of multiple epochs in compact and extended antenna configurations: we considered only the long-baseline configurations for this experiment to maintain the higher angular resolution. The data for SDP\,81 were taken during the 2014 ALMA long baseline science verification campaign, using an atypical array configuration \citep{Vlahakis2015}.

For the two SPT lenses, we used the ALMA pipeline in Common Astronomy Software Applications (CASA; \citealt{casa2022}) for the initial calibration of the data. There is an apparent spectral line emission in two spectral windows for SPT\,0532$-$50; therefore, these spectral windows are not considered further in our analysis. No line emission was detected for SPT\,0548$-$50; therefore, we use all of the spectral windows. For SDP\,81, the calibrated continuum-only dataset was obtained from ALMA via the science verification project.\footnote{\url{https://almascience.eso.org/alma-data/science-verification}} The atypical antenna configuration includes more short baselines, which results in diffuse, extended low-surface brightness emission in the image that is not useful for our analysis. Therefore, we only use baselines with lengths greater than 100~m ($\approx10^5$~k$\uplambda$) to remove sensitivity to this extended emission. For all of the datasets, the calibrated visibilities were inspected to confirm the quality of the pipeline calibration and any outlying data were flagged. No additional time or frequency averaging was done.

We performed several rounds of self-calibration to correct gain offsets between observational epochs and improve the dynamic range of the data for each target. We chose a minimum solution interval required to achieve a signal-to-noise ratio of greater than 3 for $>90$~percent of antennas, and only accepted calibrations that increase the dynamic range of the data and do not change the synthesised beam size by more than 10~percent. We iterated the procedure for as long as the dynamic range of the data improved. For SPT\,0532$-$50 and SPT\,0538$-$50, we derived gain corrections in phase and amplitude with a solution interval of each observational epoch. For SDP\,81, we derived phase-only corrections of each observational epoch (spanning approximately 50~mins each).

Deconvolved images of three lens systems are shown in Fig.~\ref{fig:clean_images}.

\subsection{Hubble Space Telescope observations}

The three lens systems were observed with the HST using the Wide-Field Camera 3 (WFC3) at 1.1 and 1.6~$\mu$m (F110W and F160W filters, respectively) as part of programmes GO-12194 (PI: Negrello) and GO-12659 (PI: Vieira). These data were obtained from the HST archive and processed using standard procedures within the {\sc astrodrizzle} package. During this process, the images were drizzled to 60~mas~pixel$^{-1}$.

\section{Lens modelling}
\label{section:modelling} 

\subsection{Mass distribution}
\label{section:smooth_model}

We parameterised the underlying global mass distribution of the lens as a singular power-law ellipsoidal density profile. Following \cite{Tessore:2015} and \cite{ORiordan:2020}, the 2D dimensionless surface mass density profile (convergence) is described by 
\begin{equation}
    \kappa(R) = \frac{3-\gamma}{2} \left ( \frac{b}{R} \right )^{\gamma -1}\, ,
\label{eq:PL}
\end{equation} where $R$ is the elliptical radius ($R^{2}=q^{2}x^{2} + y^{2}$), $\gamma$ is the 3D density slope ($\gamma=2$ is isothermal), $q$ is the axis ratio, and $b=\sqrt{q}\ R_{E}$, where $R_E$ is the Einstein radius. $\theta_q$ describes the position angle of the ellipticity, defined in degrees east of north. We also allowed for an external shear described by a strength, $\Gamma$, and angle, $\theta_\Gamma$. We labeled this model `PL'.

We then allowed flexibility in the form of multipole perturbations that add angular variations to the PL density profile to account for deviations from perfect ellipticity. These are described by the following Fourier expansion:
\begin{equation}
    \kappa_m(R,\theta) = R^{-(\gamma-1)}\left[A_m \sin(m\theta)+B_m\cos(m\theta)\right]\, ,
\label{eq:multipoles}
\end{equation} in polar co-ordinates for multipole order $m$, where $A_m$ and $B_m$ are unitless sine and cosine multipole coefficients that give the strength of the density perturbations normalised to the critical density at 1 arcsec radius from the lens centre. $\gamma$ is the slope of the density profile from Eq~(\ref{eq:PL}). We adopted multipoles of order $m=3$, which allows for asymmetrical density structure with respect to the lens centre, and order $m=4$, which can create boxy or discy shapes that are often observed in the isophotes of elliptical galaxies \citep{Bender:1988,Pasquali:2006,Hao:2006, Chaware:2014}\footnote{Multipoles of order 0 and 2 are implicit in Eq.~\ref{eq:PL}.} if aligned with the ellipticity. We consider only up to order 4 here, as these have been predicted to be sources of systematic errors in the constraints derived from lensing (e.g. \citealt{Nightingale:2022,VandeVyvere:2022,O'Riordan:2023}). We label this model component as `MP'. 

\subsection{Bayesian inference}

We employed the Bayesian pixellated lens modelling technique appropriate for interferometric data introduced by 
\citet[][see also \citealt{Vegetti:2009,Rybak:2015a,Rizzo:2018,Powell:2022}]{Powell:2021}. This method reconstructs the source emission on vertices of Delaunay triangles adapted to the lensing magnification. The source ({s}) and lens parameters ($\boldsymbol{\eta}$) were inferred by maximising the following posterior:
\begin{equation}
    P(\mathbf{s}, \boldsymbol{\eta} | \mathbf{d}) \propto P(\mathbf{d} | \mathbf{s}, \boldsymbol{\eta}) P(\mathbf{s} | {\mathbf{H}}, \lambda_{\rm s}) P(\boldsymbol{\eta})\, ,
\label{equation:posterior}
\end{equation} 
where {\bf d} is the data. $P(\mathbf{s} | {\mathbf{H}}, \lambda_{\rm s})$ is a quadratic regularising prior on the source light distribution, $\mathbf{s}$, expressed in terms of a regularisation form, ${\mathbf H}$, and strength, $\lambda_{\rm s}$. The regularisation strength is a free hyper-parameter of the model. We refer to \cite{Vegetti:2009} and \cite{Powell:2022} for more details. 

We considered three forms of regularisation: gradient, curvature, and area-weighted gradient. Gradient and curvature impose smoothness in the source surface brightness distribution by minimising the gradient or curvature between adjacent points on the Delaunay grid. The area-weighted gradient is a modification to the gradient regularisation that is weighted according to the triangle area, thereby allowing more freedom in high-magnification areas and less in regions of lower magnification. As was demonstrated by \cite{Suyu:2006}, different source surface brightness distributions may require different forms of regularisation. We adopted the form and strength of regularisation that maximises the posterior (Eq.~\ref{equation:posterior}). 

$P(\mathbf{d} | \mathbf{s}, \boldsymbol{\eta})$ is the likelihood, which we assumed to be Gaussian. As is described by \cite{Powell:2022}, it encodes all the linear operators that describe the signal propagation processes from the source to the lensed image plane and the instrumental response; that is,%
\begin{equation}
\log{P(\mathbf{d} | \mathbf{s}, \boldsymbol{\eta})} = \frac{1}{Z} e^{ -\frac{1}{2}\chi^{2} }\, ,
\end{equation} where 
\begin{equation}
    \chi^2 = \mathbf{s}_{\rm MAP}^{\rm T}\mathbf{L}^{\rm T}\mathbf{\tilde{C}}_{x}^{-1}\mathbf{L}\mathbf{s}_{\rm MAP} - 2 \mathbf{s}_{\rm MAP}^{\rm T}\mathbf{L}^{\rm T}\mathbf{d}_x + \mathbf{d}^{\rm T}\mathbf{C}^{-1}\mathbf{d}
\end{equation} and $Z = \sqrt{\rm det(2\pi\mathbf{C})}$. $\mathbf{s}_{\rm MAP}$ is the maximum a posteriori source for the lensing operator, $\mathbf{L}(\boldsymbol{\eta})$. $\mathbf{d}$ is the data, $\mathbf{d}_x$ is the naturally weighted dirty image, and $\mathbf{C}$ is the noise covariance assuming uncorrelated, Gaussian noise. Due to the observing frequencies of these data and the antenna size of ALMA, the antenna response becomes significant just a few arcseconds from the pointing centre (which in our case is also the phase centre). We accounted for the loss of sensitivity using a Gaussian function with a half-power width of $1.13\times\lambda/D$, where $\lambda$ is the central frequency and $D$ is the antenna diameter (12~m).\footnote{ALMA Cycle 7 Technical Handbook \url{https://arc.iram.fr/documents/cycle7/ALMA_Cycle7_Technical_Handbook.pdf}} We implemented this as a diagonal operator, $\mathbf{P}$, such that the image-plane noise covariance takes the form $\mathbf{\tilde{C}}_{x}^{-1} = \mathbf{P}^{\rm T}\mathbf{D}^{\rm T}\mathbf{C}^{-1}\mathbf{D}\mathbf{P}$, where $\mathbf{D}$ is a discrete Fourier transform (the instrumental response). The terms $\mathbf{D}^{\rm T}\mathbf{C}^{-1}\mathbf{D}$ together are equivalent to a convolution with the dirty beam, which was performed using a fast Fourier transform.

The posterior probability distributions of the lens parameters, $\boldsymbol{\eta}$, source surface brightness, and source regularisation were inferred using {\sc MultiNest} \citep{Feroz:2009,Feroz:2019}. The Bayesian evidence was computed from the integral
\begin{equation}
    \mathcal{E} = \int {P(\mathbf{d} | \mathbf{s}, \boldsymbol{\eta})}\, {P(\mathbf{s}, \lambda_s)}\, {P(\boldsymbol{\eta})}\, \delta\lambda_s\, \delta\mathbf{s}\, \delta\boldsymbol{\eta}\, ,
\end{equation} 
using 240 live points and importance sampling with a constant efficiency of 0.05 to improve the accuracy of the evidence calculation \citep{Feroz:2019}. From tests on mock data, we have found that a minimum of 200 live points are required to recover the correct posterior for these data and the number of model parameters explored here (we note that this may be data-dependent).

Following an initial optimisation (see \citealt{Vegetti:2009} for details), we set uniform priors for the lens model parameters at $\pm10$ to 20~percent of the optimised values, except the external shear, which was uniform in log space ($\log\Gamma\in[-3,-1]$) with a free position angle ($\theta_\Gamma\in[-90,90]$ deg). We marginalised over the source surface brightness and left the regularisation strength as a free parameter, which was uniform in log space.

For the multipole coefficients, we set uniform priors of $\pm0.01$. These priors are motivated by observations of isophotes of elliptical galaxies by \cite{Hao:2006} and the analysis of simulations by \citet[][see also \citealt{Powell:2022}]{Kochanek:2004}. Additionally, larger $m=4$ amplitudes paired with low ellipticity could produce exotic lensing configurations with six or more images \citep{Evans:2001,Hao:2006} and we are not aware of any such configurations reported for galaxy-scale lenses. We compared the Bayesian evidence from MultiNest for the different regularisation types to determine the appropriate type for the remainder of the analysis. We find that gradient regularisation gives the strongest evidence for SDP\,81 and SPT\,0538$-$50, while the area-weighted gradient gives stronger evidence for SPT\,0532$-$50.

\begin{figure}
    \centering
         \includegraphics[width=0.49\textwidth]{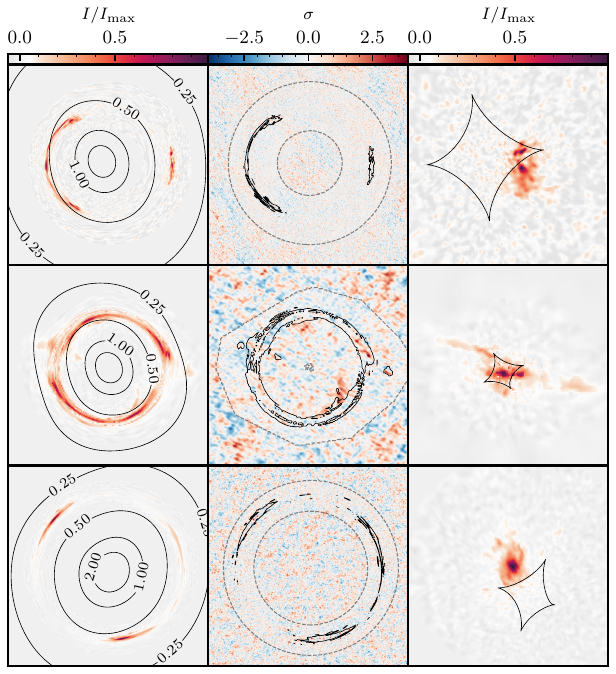}
         \caption{Maximum a posteriori PL+MP lens models. Rows, from top to bottom: SDP\,81, SPT\,0532$-$50, and SPT\,0538$-$50. First column: Model lensed surface brightness distribution, with contours of the convergence in units of the critical surface-mass density in black. 
         Second column: Noise-normalised residuals (data$-$model); black contours show the model lensed surface brightness distribution; dashed grey contours show the mask. Third column: Model source surface brightness distribution, with caustics in black. The colour scale is the same for each column.}
         \label{fig:lens_models}  
\end{figure}

\begin{figure}
    \centering
        \includegraphics[width=0.98\columnwidth]{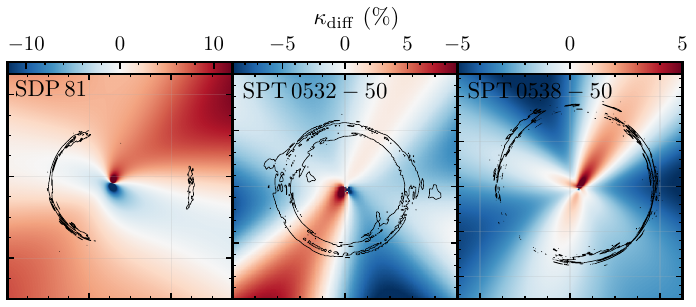}%\vspace{-8pt}
         \caption{Difference between PL+MP and PL maximum a posteriori model convergences (i.e. $\kappa_{\rm diff} = (\kappa_{\rm PL+MP}-\kappa_{\rm PL}) / \kappa_{\rm PL}$). The black contours are the best model lensed surface brightness distribution. }
         \label{fig:kappa_diff}
\end{figure}

\begin{table*}
    \caption{Posterior distributions of the parameters of lens models. The uncertainties given are the weighted 1st and 99th percentile ranges of the marginalised posterior sampling with MultiNest. Positions are given relative to the observation phase centre, given in Table~\ref{table:obs}. The Bayes factor ($\mathcal{K}$) is relative to the PL model. All angles are defined east of north.}\vspace{-24pt}
    \centering
    \setlength{\tabcolsep}{3pt}
    \renewcommand*{\arraystretch}{1.5}
    \adjustbox{width=\textwidth}{
    \begin{tabular}{l p{0.5cm} C{2.5 cm}  C{2.5 cm}  p{0.4cm}  C{2.5 cm}  C{2.5 cm}  p{0.4cm}  C{2.5 cm}  C{2.5 cm} } \\
    % \begin{tabular}{l l c  c  l  c  c  l  c  c } \\
    && \multicolumn{2}{c}{SDP\,81} && \multicolumn{2}{c}{SPT\,0532$-$50} && \multicolumn{2}{c}{SPT\,0538$-$50} \\ \cline{3-4} \cline{6-7} \cline{9-10}  
    &&  PL & PL+MP && PL & PL+MP && PL & PL+MP \\ \hline 
    $x_l$           && $0.542^{+0.003}_{-0.011}$ & $0.560^{+0.003}_{-0.003}$ && $2.1043^{+0.0003}_{-0.0007}$ & $2.1003^{+0.0004}_{-0.0003}$ && $0.1896^{+0.0008}_{-0.0008}$ & $0.1907^{+0.0007}_{-0.0026}$  \\ 
    $y_l$           && $-0.170^{+0.005}_{-0.009}$ & $-0.149^{+0.003}_{-0.004}$ && $1.8706^{+0.0001}_{-0.0003}$ & $1.8700^{+0.0004}_{-0.0003}$ && $-0.0277^{+0.0009}_{-0.0016}$ & $-0.0254^{+0.0013}_{-0.0011}$  \\ 
    $R_E$      && $1.609^{+0.007}_{-0.002}$ & $1.611^{+0.003}_{-0.002}$ && $0.5420^{+0.0001}_{-0.0001}$ & $0.5420^{+0.0002}_{-0.0002}$ && $1.7237^{+0.0003}_{-0.0010}$ & $1.7243^{+0.0004}_{-0.0004}$  \\ 
    $q$             && $0.794^{+0.010}_{-0.031}$ & $0.832^{+0.009}_{-0.008}$ && $0.816^{+0.005}_{-0.008}$ & $0.840^{+0.005}_{-0.005}$ && $0.894^{+0.008}_{-0.003}$ & $0.876^{+0.004}_{-0.004}$  \\ 
    $\theta_q$      && $12^{+4}_{-2}$ & $ 6^{+2}_{-1}$ && $28.3^{+1.3}_{-0.3}$ & $23.6^{+0.5}_{-0.6}$ && $152^{+1}_{-1}$ & $153^{+1}_{-1}$  \\ 
    $\gamma$        && $1.97^{+0.04}_{-0.13}$ & $2.00^{+0.03}_{-0.07}$ && $2.19^{+0.04}_{-0.03}$ & $2.20^{+0.04}_{-0.02}$ && $2.22^{+0.03}_{-0.06}$ & $2.23^{+0.02}_{-0.03}$  \\ 
    $\Gamma$        && $0.032^{+0.005}_{-0.019}$ & $0.037^{+0.004}_{-0.007}$ && $0.015^{+0.002}_{-0.002}$ & $0.022^{+0.003}_{-0.001}$ && $0.012^{+0.001}_{-0.001}$& $0.011^{+0.001}_{-0.001}$ \\ 
    $\theta_\Gamma$ && $-8^{+4}_{-12}$ & $ 8^{+2}_{-2}$ && $13^{+2}_{-6}$ & $26^{+1}_{-1}$ && $15^{+3}_{-4}$ & $19^{+2}_{-2}$ \\ 
    $A_3$           && - & $0.0018^{+0.0009}_{-0.0009}$ && - & $0.0047^{+0.0003}_{-0.0004}$ && - & $-0.0078^{+0.0006}_{-0.0013}$ \\ 
    $B_3$           && - & $0.0034^{+0.0006}_{-0.0007}$ && - & $-0.0019^{+0.0003}_{-0.0003}$ && - & $0.0049^{+0.0022}_{-0.0009}$ \\ 
    $A_4$           && - & $-0.0032^{+0.0015}_{-0.0013}$ && - & $-0.0037^{+0.0006}_{-0.0007}$ && - & $-0.0033^{+0.0012}_{-0.0009}$ \\ 
    $B_4$           && - & $0.0041^{+0.0016}_{-0.0014}$ && - & $-0.0060^{+0.0007}_{-0.0007}$ && - & $-0.0078^{+0.0014}_{-0.0021}$ \\ \hline 
    $\mathcal{K}$   && $\equiv0$ & 28 && $\equiv0$ & 75 && $\equiv0$ & 120 \\ \hline
    \end{tabular} }
\label{table:nested}
\end{table*}

\section{Results}
\label{section:results}

\subsection{Evidence for complex angular structure}

Table~\ref{table:nested} shows the marginalised posterior of the lens parameters and the relative evidence for the three datasets and the two different lens models tested here. The maximum a posteriori PL+MP lens model is shown in Fig.~\ref{fig:lens_models} for each system. The differences in the convergence between the lens models are shown in Fig.~\ref{fig:kappa_diff} and have convergence differences up to $\approx5$ to 10~percent at the Einstein radius.

\begin{figure}
    \centering
    \includegraphics[width=\columnwidth]{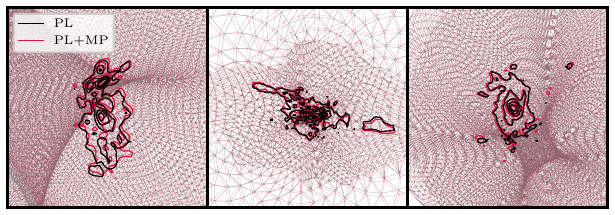}
    \caption{Contours of the sources and source grids for the maximum a posteriori lens models. Left to right: SDP\,81, SPT\,0532$-$50, and SPT\,0538$-$50. There are only minor visual differences in the source structure between lens models.}
    \label{fig:sources}
\end{figure}

The sources are not visually different between the PL and PL+MP models, as can be seen in Fig.\ref{fig:sources}. Nevertheless, the PL+MP is preferred over a PL for all three lenses, with a relative Bayes factor (difference in $\ln{\mathcal{E}}$) of 28, 75, and 120 for SDP\,81, SPT\,0532$-$50 and SPT\,0538$-$50, respectively (i.e. strong to decisive evidence in favour of more complex angular structure; \citealt{Kass:1995}). In all cases, the improvement in model evidence is dominated by the increase in source regularisation rather than a change in the residuals (i.e. $\chi^{2}$). This indicates that the models can fit the data equally well, but the source can be smoother in the PL+MP case. This is consistent with the findings of \cite{Powell:2022}, who compared various mass models for a system with a lensed radio jet. We interpret this as the regularisation trying to enforce the lens equation (i.e. better focus the source), as is discussed in Section~6.3 of \cite{Powell:2022}.

We find non-zero $m=3$ and $m=4$ amplitudes of $\approx0.01$ for all three lenses, suggesting a non-negligible departure from the PL model. The model parameters of the PL fitted with and without multipoles are generally in agreement for each of the three lens systems, although some parameters deviate by several sigma. Some notable differences are the positions of the mass centres for each system, the axis ratios, and position angles. Interestingly, the shear strength is largely consistent between the PL and PL+MP models, although the shear position angle does differ by several sigma for SDP\,81 and SPT\,0532. Altogether, we do not see clear evidence that the PL model is not compensating for some unmodelled angular structure when multipoles are not included, in contrast with the findings of \cite{Etherington:2023}.

We find that the PL+MP model posterior for SDP\,81 differs significantly in ellipticity, external shear, and multipole amplitudes from that of \cite{Hezaveh:2016} using the same set of observations. The underlying PL model is similar to those of previous studies using these visibility data from baselines $<2$~km \citep{Rybak:2015a} and deconvolved images \citep{Dye:2015,Inoue:2016,Tamura:2015,Wong:2015}. We were not able to find a good model for the data using the PL+MP parameters reported by \cite{Hezaveh:2016}. 

\subsection{Isophote and isodensity shapes}

Anisotropic shapes are a known feature of the stellar distribution of elliptical galaxies. Since stars compose the vast majority of the baryonic matter of elliptical galaxies, by measuring stellar isophotes for these lens galaxies we can test to what extent the mass-density distribution from lensing may be shaped by baryons. 

For SDP\,81, there is lensed emission detected in the F160W HST/WFC3 imaging that is blended with the lens light \citep{Rybak:2015b}. We modelled this lensed emission with the PL lens model (Table~\ref{table:nested}) while simultaneously fitting the lens galaxy light (see \citealt{Ritondale:2019a} for details of this methodology). We then subtracted model lensed emission from the data so as not to include it in our subsequent isophote fitting. After masking the light from any interposing objects, we used the {\sc isophote} tool in the Python package {\sc photutils} \citep{Bradley:2023} that uses an iterative method described by \cite{Jedrzejewski:1987} to fit elliptical isophotes to the lens galaxy light. These Fourier modes have the function
\begin{equation}
    R_m(\theta) = R_0 + a_m \sin(m\theta)+ b_m\cos(m\theta) 
\end{equation}
in polar coordinates, where $m$ is the harmonic mode ($m=3,4$) and $R_0$ is the elliptical path for that isophote at the angles defined by $\theta$. It should be noted that the isodensity multipole parameters are not the same as those of the lens model parameters because the multipoles in the lens model are spherical and the elliptical power-law model induces a radial dependence. We converted the multipole shape parameters to an amplitude and position angle for ease of interpretation; the amplitude of a Fourier mode is described by
\begin{equation}
    \eta_m = \sqrt{a_m^2 + b_m^2}\, ,
\end{equation} and the position angle is described by
\begin{equation}
    \phi_m = \frac{1}{m} \arctan{\frac{b_m}{a_m}}\, .
\end{equation} 
Boxy or discy shapes can be discerned from the difference between $\eta_4$ and the ellipticity position angle ($\theta_q$), where 45, 135~deg is a truly boxy shape and 0, 90~deg is a truly discy shape.

\begin{figure}
    \centering
    \includegraphics[width=\columnwidth]{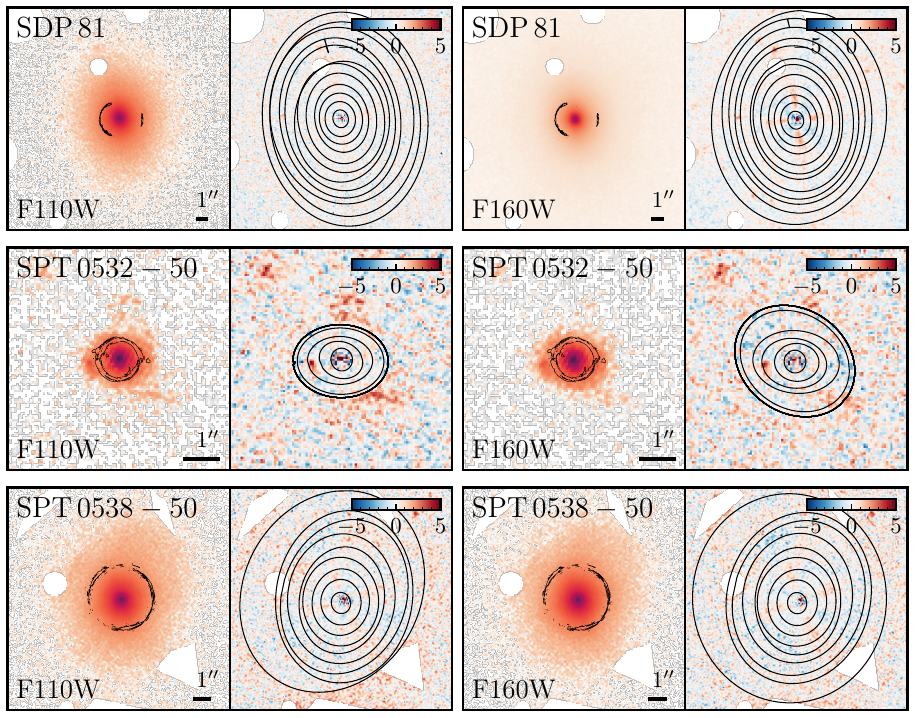}
    \caption{Best-fit isophotes for the three lenses in HST/WFC3 F110W and F160W filters. Left: Data (log scale) and ALMA contours in black. Neighbouring objects have been masked. Right: Noise-normalised residuals (data$-$model) with a sub-sample of isophote shapes.}
    \label{fig:isophotes}
\end{figure}

Fig.~\ref{fig:isophotes} shows the isophote fits to the F110W and F160W HST/WFC3 imaging of the three lens galaxies. F110W and F160W have similar isophote fits. However, SPT\,0532$-$50 has the most scatter and the least well-constrained isophotes because the lens is significantly fainter than those of the other two lens systems. 

\begin{figure}
    \centering
    \includegraphics[width=\columnwidth]{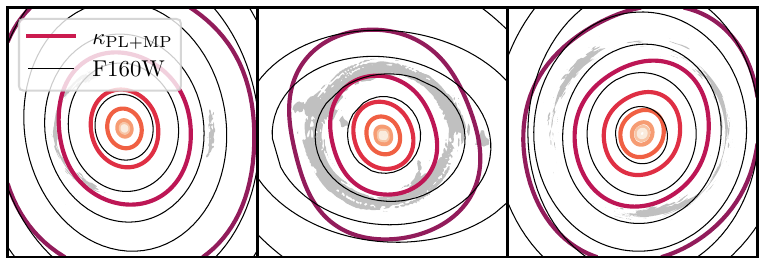}
    \caption{Comparison of the total convergence of the maximum a posteriori PL+MP lens model ($\kappa_{\rm PL+MP}$; colour) and a sub-sample of F160W isophotes (black). Left to right: SDP\,81, SPT\,0532$-$50, and SPT\,0538$-$50.} 
    \label{fig:converence_isophotes}
\end{figure}

\begin{figure*}
    \centering
    \includegraphics[width=0.98\textwidth]{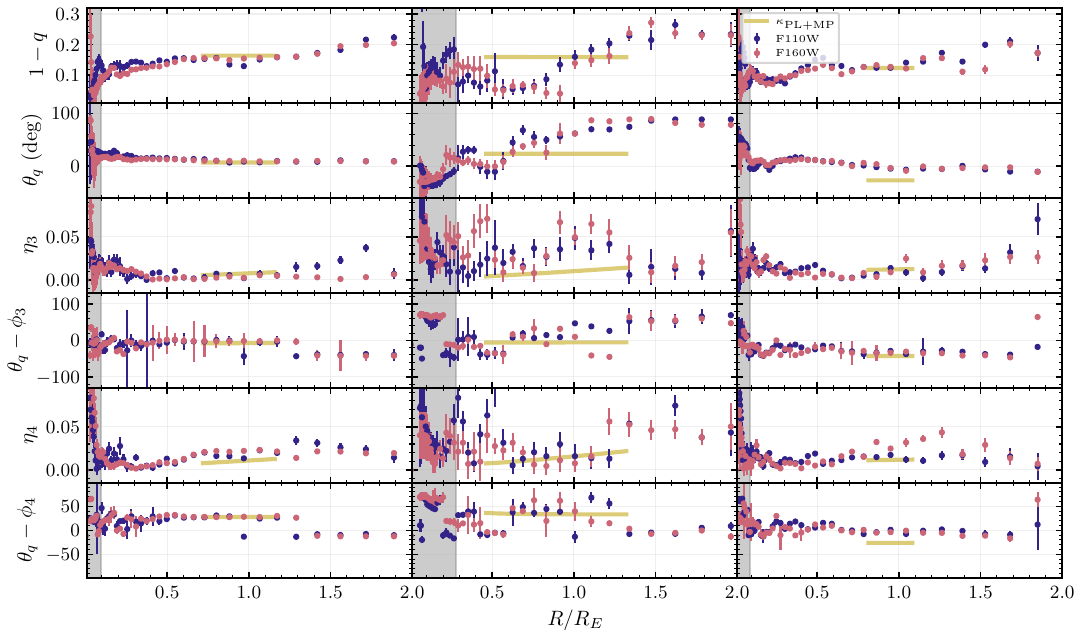}
    \caption{Shape parameters against semi-major axis radius as a function of the Einstein radius. Columns, from left to right: SDP\,81, SPT\,0532$-$50, and SPT\,0538$-$50. The points show the F110W (blue) and F160W (mauve) isophote fits and their $1\sigma$ errors. The yellow curves show the isodensity parameters for PL+MP for the approximate range of radii where there is lensed emission (the convergence $1\sigma$ errors are too small to see). The shaded grey region denotes the FWHM of the F160W point spread function (the larger of the two filters). All angle definitions are east of north. The multipole symmetry means that $\phi_3$ cycles over 120 deg and $\phi_4$ over 90 deg, so jumps of those magnitudes are not physical.}
    \label{fig:isophote_params}
\end{figure*}

A comparison of the total lens model convergence (PL+MP; isodensity) and the F160W isophotes is shown in Fig.~\ref{fig:converence_isophotes}. In addition, we show comparisons of the isophote shape parameters and isodensity parameters from lensing as a function of galaxy radius for the three lens systems in Fig.~\ref{fig:isophote_params}. 

None of the lensing galaxies are truly boxy or discy in their isodensity shapes, but rather somewhere in between. However, those of SPT\,0532$-$50 appear quite boxy (Fig.~\ref{fig:converence_isophotes}). SDP\,81 is largely consistent in terms of its isophote and isodensity shapes, while SPT\,0532$-$50 and SPT\,0548$-$50 show disagreement. For SPT\,0532$-$50, the isophote ellipticity position angle is offset by about 40 deg relative to the isodensity profile at the Einstein radius, and for SPT\,0538$-$50 this offset is around 25 deg. 

In all cases, the ellipticities ($1-q$) are in good agreement at the Einstein radius. We also see that the isophotes for all three lens galaxies have some change in ellipticity and ellipticity position angle with radius, which is not accounted for in the lens model. We tested whether the position angles of the isophotes could be used to describe the mass distribution for SPT\,0532$-$50 and SPT\,0538$-$50, given the multipole flexibility. This was done by fixing the ellipticity position angle to that of the isophotes and leaving all other parameters free to optimise. The best fits result in significant residual emission and a more disrupted source, confirming that the isophote ellipticity position angles are indeed inconsistent with those of the total mass distribution. 

\section{Discussion}
\label{section:discussion}

\subsection{Origin of the angular structure}

The three lenses in our sample are massive elliptical galaxies, based on their optical colours and morphologies. Massive elliptical galaxies are generally slow, anisotropic rotators characterised by triaxial shapes, and are frequently found to have boxy or discy isophote morphologies to their rest-frame optical stellar emission \citep{Bender:1988,Cappellari:2016}. These features are likely reflected to some extent in their isodensity structure. Therefore, one might assume that this could explain the low-amplitude multipole components that we detect in our lensing analysis if the total mass follows that of the stellar morphology. 
 
We find isodensity multipole amplitudes in the range of 0.1 to 1~percent, which is comparable to what is typically seen in isophotes of elliptical galaxies \citep{Kormendy:1989,Kormendy:2009}, and similar to what has been explored in recent theoretical work \citep{O'Riordan:2023,VandeVyvere:2022,Cohen:2024}. However, when compared directly with these galaxies, there are minor differences in isodensity and isophote shapes, suggesting that their stellar distributions are not completely consistent with those of the total projected mass distributions. As the infrared stellar emission likely traces the vast majority of the baryonic mass in the inner halo, any differences could be caused by a different shape to the galaxy's dark matter halo that can contribute significantly to the density at the Einstein radius \citep{Auger:2010,Sonnenfeld:2012,Oldham:2018}. 

Numerical simulations predict that the shape of a dark matter halo is influenced by baryonic structures and its merger history \citep{Prada:2019,Chua:2019}. As a result, dark matter haloes can exhibit ellipticity, twists, and misalignment with baryons that increase with radius \citep{Liao:2017,Emami:2021,Han:2023}. However, these simulations do not resolve the inner radii probed by galaxy-scale lensing, nor do they resolve them well, so they do not make clear predictions for our data.

The lens galaxies in this work may be part of groups in which the halo is disturbed by ongoing mergers and interactions: SPT\,0538$-$50 and SDP\,81 have several close galaxies in projection in the HST imaging (these have been masked in Fig.~\ref{fig:isophotes}) that suggest the lens is the central galaxy of a group (we note that the redshifts of any neighbours are not reported in catalogues, so it is not necessarily clear what they are or whether they are associated). On the other hand, this does not appear to be the case for SPT\,0532$-$50, which exhibits the strongest misalignment between its stellar and total mass distribution while being seemingly isolated. The lens of SPT\,0532$-$50 is significantly less massive than the others and has a higher redshift, so it may be in a different stage of its evolution \citep{Despali:2014,Cataldi:2023} or follow a different formation pathway \citep{Penoyre:2017,Lagos:2018}.

Alternatively, rather than the halo of the lens itself, the multipoles could be caused by density perturbations extrinsic to the lens galaxy. Galaxies at a different redshift close in projection could also be perturbing the lensed images. Here, the distributions of these nearby galaxies do not appear to correlate with the ellipticity position angle from lensing: for example, SPT\,0532$-$50 demonstrates the largest offset, but no nearby galaxies are apparent in the infrared imaging. However, these perturbers need not be directly observable; \cite{O'Riordan:2023} find that the anisotropic distribution of dark sub-haloes may be interpreted as third and fourth-order multipoles in the lensing convergence. Line-of-sight interloper haloes may also induce multipole structure, as they produce a similar lensing effect and are expected to contribute significantly to the lensing perturbations \citep{Despali:2018,Amorisco:2022}. Such a scenario could be occurring for SDP\,81 and SPT\,0538$-$50, where the $m=4$ amplitudes are inconsistent with those of the light. Additionally, SPT\,0532$-$50, where the total mass shows the largest offset from the light, is also at the highest lens redshift, where there will be more lensing signal from low-mass haloes along the line of sight \citep{Despali:2018}.

While there are several possible contributions to the inconsistency between the isodensity and isophotal shapes for these lens galaxies, it may be possible to discern these contributions with non-analytical lens modelling methods  (e.g. \citealt{Vegetti:2009,Vernardos:2022,Galan:2022}) if different mass components produce different lensing signatures. For example, \cite{Galan:2022} used a technique involving wavelets to demonstrate that it may be possible to differentiate the lensing signal induced by a population of low-mass haloes from an intrinsic multipole structure in the lensing galaxy, which could constrain dark matter models. Additionally, intrinsic multipole shapes or the lack thereof in the lens galaxies themselves may provide a test of self-interacting dark matter \citep[e.g.][]{Brinckmann:2018}, provided that the baryons do not significantly affect the shape of haloes in the regions of interest \citep{Despali:2022b}.

\subsection{Consequences for lensing studies}

Complex angular structure could have significant effects on lensing observables that are relevant for all lines of research in gravitational lensing, particularly measuring the Hubble constant from time delays (e.g. \citealt{Wong:2020}) and constraints on dark matter models via the detection of low-mass dark matter haloes (e.g. \citealt{Ritondale:2019b,Gilman:2020a}). Recent works have attempted to address this by considering composite models consisting of the observed stellar component (with a constant mass-to-light ratio) and an elliptical dark matter profile (e.g. \citealt{Rusu:2020,Nightingale:2022,Chen:2022}), or explored the consequences of boxy and discy shapes \citep{VandeVyvere:2022}. However, composite lens models, even when the two components are allowed to be misaligned, only allow for additional angular complexity of order $m=2$. Studies that consider the consequences of higher-order angular structure on lensing observables typically consider $m=4$ and not $m=3$ \citep{VandeVyvere:2022,Nightingale:2022,Gilman:2023}, and often limit $m=4$ structure to true boxy or discy shapes (i.e. $\theta_q - \phi_4 = 0$). We have shown here that these models do not encompass the true angular complexity of lensing galaxies. Future work involving a larger sample of lenses is needed to provide statistics on the level of complex angular structure in lensing galaxies. Such data will provide an essential test of systematic errors on cosmological constraints derived from strong gravitational lensing.

\section{Conclusions}
\label{section:conclusions}

The singular ellipsoidal power-law model has been the predominant model in galaxy-scale strong lensing studies for decades. However, as observatories reach higher angular resolutions, the data becomes sensitive to deviations from this simple assumption. We have shown that significant multipole expansions of the power-law model up to the fourth order with Bayes factors between 28 and 120 can be measured with ALMA observations at an angular resolution of $\approx$\,30\,mas. While this is significantly lower evidence than for very-long-baseline interferometry (VLBI) data at an angular resolution of $\sim$mas \citep{Powell:2022}, the current relative abundance of lensed dusty star-forming galaxies compared to extended arcs in VLBI observations means that future observations with ALMA could provide a statistically meaningful sample of lensing angular structure. Additionally, future surveys, such as with \textit{Euclid} and the Square Kilometre Array, will allow for the selection of a large sample of useful targets for high-resolution interferometry \citep{Serjeant:2014,McKean:2015}.

Previous constraints on dark matter models via gravitationally lensed arcs have relied on rare individual detections of low-mass perturbers to constrain the shape of the dark matter halo mass function \citep{Vegetti:2014,Vegetti:2018,Ritondale:2019a,Enzi:2021}. Searches for such perturbers could be susceptible to systematic biases if the angular structure in the lens is not accounted for \citep{Nightingale:2022}. Indeed, \cite{O'Riordan:2023} used mock observations to show that angular structure in the lens could create false detections in HST-like data. Here we find, using data of a higher quality than those provided by the HST, that angular structure with comparable amplitudes does exist in such systems, and can be accounted for using multipoles. The differences we find between the lensing angular structure and stellar isophotes suggest a contribution from non-baryonic mass structure, which could be caused by a misalignment of the galaxy's dark matter halo \citep{Liao:2017} and/or the influence of a population of low-mass perturbers \citep{O'Riordan:2023}. Our future work will use pixellated potential corrections to shed further light on the nature of the complex angular structure identified for the sample of lenses presented in this paper.

\begin{acknowledgements}
HRS, DMP and SV acknowledge funding from the European Research Council (ERC) under the European Union's Horizon 2020 research and innovation programme (LEDA: grant agreement No. 758853). JPM and DW acknowledge support from the Netherlands Organization for Scientific Research (NWO) (Project No. 629.001.023) and the Chinese Academy of Sciences (CAS) (Project No. 114A11KYSB20170054). This work is based on the research supported in part by the National Research Foundation of South Africa (Grant Number: 128943).
Our analysis made use of SciPy, NumPy, Matplotlib, Astropy and Photutils packages for Python \citep{Virtanen:2020,Harris:2020,Hunter:2007,Astropy:2013,Astropy:2018,Bradley:2023}. Figs.~\ref{fig:clean_images}, \ref{fig:lens_models}, \ref{fig:isophotes} colour scheme courtesy of Rebecca Levy following \cite{Levy:2021}. We used ALMA data with project codes 2011.0.00016.SV and 2016.1.01374.S. ALMA is a partnership of ESO (representing its member states), NSF (USA) and NINS (Japan), together with NRC (Canada), MOST and ASIAA (Taiwan), and KASI (Republic of Korea), in cooperation with the Republic of Chile. The Joint ALMA Observatory is operated by ESO, AUI/NRAO and NAOJ. We used observations made with the NASA/ESA \textit{Hubble} Space Telescope obtained from the Space Telescope Science Institute, which is operated by the Association of Universities for Research in Astronomy, Inc., under NASA contract NAS 5–26555. These observations are associated with proposal IDs 12194 and 12659.
\end{acknowledgements}

% \section*{Data Availability}

% All observations reported in this work are publicly available in the ALMA archive (\url{https://almascience.eso.org/aq}) and HST archive (\url{https://mast.stsci.edu/search/ui/#/hst}). The datasets generated during this study are available from the corresponding author upon reasonable request.

% \section*{Code Availability}

% The lens modelling methodology used in this work is detailed in \cite{Powell:2021} (see also \citealt{Vegetti:2009,Rybak:2015a,Rizzo:2018,Ritondale:2019a}). The reader interested in using this software should contact SV. All other software used in this work is publicly available.

%%%%%%%%%%%%%%%%%%%%%%%%%%%%%%%%%%%%%%%%%%%%%%%%%%

%%%%%%%%%%%%%%%%%%%% REFERENCES %%%%%%%%%%%%%%%%%%

\bibliographystyle{aa}
\bibliography{references} 

\end{document}